\documentclass[aps,prl,preprint,groupedaddress]{revtex4}
\usepackage{color}
\usepackage[dvips]{graphicx}

\begin{document}


\title{Multipolar expansion of orbital angular momentum modes}



\author{Gabriel Molina-Terriza}
\email[]{gabriel.molina@icfo.es}

\affiliation{ICFO-Institut de Ciencies Fotoniques, Parc Mediterrani
de la Tecnologia, 08860, Castelldefels, Barcelona, SPAIN}
\affiliation{ICREA- Instituci\'{o} Catalana de Recerca i Estudis
Avan\c{c}ats, 08010, Barcelona, Spain}

\date{\today}

\begin{abstract}
In this letter a general method for expanding paraxial beams into
multipolar electromagnetic fields is presented. This method is
applied to the expansion of paraxial modes with orbital angular
momentum (OAM), showing how the paraxial OAM is related to the
general angular momentum of an electromagnetic wave. This method can
be extended to quasi-paraxial beams, i.e. highly focused laser
beams. Some applications to the control of electronic transitions in
atoms are discussed.
\end{abstract}


\maketitle

The angular momentum of light fields continues to draw the interest
of the scientific community. Since the work of Allen and coworkers,
showing that the angular momentum components of a paraxial light
beam, i.e. the spin and the orbital angular momentum (OAM), could be
controlled separately \cite{allen-book}, there has been an ongoing
discussion regarding concepts as the conservation of angular
momentum in different light interaction processes
\cite{bloembergen1,berzanskis97,barbosa3}, the angular momentum flux
\cite{barnett02} or the measurement of the angular momentum of light
\cite{leach1,vasnetsov03}.

In parallel to this scientific discussion, there has been an
explosion of applications of the angular momentum of light, and in
particular to the orbital part, covering areas as diverse as
trapping and rotating microparticles \cite{grier}, astrophysical
measurements \cite{harwit, grover, thidePRL07} or quantum
information \cite{molinareview}. Also, recently there has been a
renewed interest in the interaction of the orbital angular momentum
of light with atomic ensembles \cite{inoue1} and the electronic
structure of atoms and molecules \cite{alexandrescu06}.

On the other hand, a general electromagnetic field it is known to
have an angular momentum content \cite{Jackson}. One can even
construct solutions to the electromagnetic field with the right
spherical symmetries to have well defined values of the total
angular momentum and one component of the angular momentum. This
kind of solutions receive the general name of multipolar solutions
of the electromagnetic field. Up to now, the multipolar fields and
the paraxial modes with OAM have been used separately in different
regimes. In this letter I try to close this gap and I calculate the
multipolar content of any paraxial beam and in particular of the OAM
modes.

A general electromagnetic field in free space has two components of
the angular momentum (${\mathbf J}$): the spin part (hereby denoted
by ${\mathbf S}$), which is related to the vectorial character of
the field, and the orbital angular momentum (OAM) ${\mathbf L}$,
which is related to the spatial structure of the field
\cite{Jackson}. Both parts are needed to generate rotations in
space, and consequently only the combination of the two components
plays a meaningful role in the rotational symmetries of an
electromagnetic wave, i.e. ${\mathbf J}={\mathbf L}+{\mathbf S}$
\cite{RoseAM}. Multipolar modes are precisely a set of solutions to
the Maxwell equations which are eigenvectors of the square of the
total angular momentum $J^2$, one component of the total angular
momentum, i.e. the $z$-component, $J_z=L_z+S_z$ and the parity
operator. The exact form of these fields can be found for example in
\cite{RoseAM}. In this letter, I will just use the properties of the
vector potential of the multipolar monocromatic fields in the
solenoidal gauge. Then following ref. \cite{RoseAM} the vector
potentials of the set of multipolar fields is ${\mathbf
A}_{jm}^{(x)}$, where $x=m$ represents the magnetic multipoles and
$x=e$ the electric multipoles. Both classes of multipoles have
eigenvalues ${\mathbf J}^2{\mathbf A}_{jm}^{(x)}=j(j+1){\mathbf
A}_{jm}^{(x)}$ and $J_z{\mathbf A}_{jm}^{(x)}=m{\mathbf
A}_{jm}^{(x)}$, but magnetic and electric multipoles differ in their
parity. Any given electromagnetic field can be fully expanded onto
multipolar fields, if we also add a longitudinal field of the form
${\mathbf A}_{jm}^{(l)}={\mathbf \nabla} \xi_j Y_{jm}$, with
$Y_{jm}$ being the usual spherical harmonics.


Let me start by constructing an electromagnetic field by superposing
rotated circularly polarized plane waves:
\begin{equation}
{\mathbf A}=\int_0^\pi\sin(\theta)d\theta\int_0^{2\pi}d\varphi
g(\theta,\varphi) R(\theta,\varphi) {\mathbf e}_p \exp(ikz)
\label{rotation}
\end{equation}
where the original plane wave propagates along the $z$ direction and
is right (left) handed polarized when $p=1$ ($p=-1$). The altitude
and azimuthal rotation angles are given respectively by
$(\theta,\varphi)$. The operator $R$, rotates the vector field in
the usual manner: $R(\theta,\varphi) {\mathbf e}_p
\exp(ikz)=M^{-1}{\mathbf e}_p\exp(i k \hat{{\mathbf z}}\cdot
(M{\mathbf r}))$. Note that this field is completely general and
does not have to fulfil paraxiality. In particular, it can be used
to describe beams highly focused with aplanatic lenses
\cite{novotnyhecht}. The function $g(\theta,\varphi)$ controls the
relative amplitudes of the plane waves.

The main idea of this letter is to find a family of fields using
suitable functions $g(\theta,\varphi)$, which in the limit of
paraxiality can be identified with the usual orbital angular
momentum modes. Let me then develop the rotation operator to the
lowest orders in the angle $\theta$ to obtain
\begin{equation}
R(\theta,\varphi) {\mathbf e}_p e^{ikz}\approx e^{i k \theta \rho
\cos (\phi-\varphi)+ i k z - i \frac{(k\theta)^2}{2k}z} (e^{-i p
\varphi}{\mathbf e}_p-\frac{\theta}{\sqrt{2}}\hat{{\mathbf z}}),
\label{loworder}
\end{equation}
where I use cylindrical coordinates $(\rho,\phi,z)$. Now, compare
this result with a circularly polarized paraxial field of the kind
${\mathbf A}={\mathfrak F}_l(\rho,z)\exp(i k_z z)\exp(i l \phi)
{\mathbf e}_p$. Any paraxial beam can be expressed as a convenient
superposition of this kind of modes \cite{Molina01}, so our results
apply to any paraxial beam. Besides, this kind of beams are
cylindrically symmetric and thus are eigenmodes of both the OAM and
the spin AM components in the $z$ direction, i.e. $L_z$ and $S_z$.
Note, however, that this feature is in principle only valid in the
paraxial approximation, when $k_z\approx k$ and $F_l$ is slowly
varying in $z$ with respect to $1/k$. This paraxial field can be
reexpressed in its Fourier components as ${\mathbf A}=\int_0^\infty
k_r\, d k_r \int_0^{2\pi} d\phi_k f_l(k_r)\exp(i l \phi_k) \exp(i
k_r r cos(\phi-\phi_k)) \exp(i k z-i k_r^2/(2k) z) {\mathbf e}_p$.

It is clear from inspection of Eqs. (\ref{rotation}) and
(\ref{loworder}) that if we do the correspondence
\begin{eqnarray}
&&k_r=k \sin(\theta), \; \phi_k=\varphi, \nonumber \\
&&g(k_r/k,\phi_k) =k^2\, f_l(k_r) \exp\left(i (l+p) \phi_k\right),
\label{changevar}
\end{eqnarray}
then the two expansions of the paraxial beam, i.e. in Fourier
components and in rotated plane waves Eq. (\ref{rotation}), are one
and the same, within the paraxial approximation. Then, the set of
functions $g$, fulfilling Eq. (\ref{changevar} defines the
sought-after family of fields. Note that, in general, we can use
functions $f_l(k_r)$ which are not paraxial, thus defining
cylindrically symmetric electromagnetic fields.

This identification is useful, because now I will use the expansion
of a rotated plane wave into multipole waves
\cite{RoseAM,Rosemultipole} and perform the integral over the
rotation parameters. The expansion reads
\begin{eqnarray}
R(\theta,\varphi){\mathbf e}_p
e^{ikz}&=&\sum_{j=1}^{\infty}\sum_{m=-j}^j i^j (2j+1)^{1/2}
\nonumber
\\ && D_{m\,p}^j(\varphi,\theta,0)[ {\mathbf A}_{jm}^{(m)} + i p {\mathbf
A}_{jm}^{(e)}], \label{pwaverot}
\end{eqnarray}
where $D_{m\,p}^j(\varphi,\theta,0)$ are the matrices of rotation
for irreducible tensors of order $j$. This matrix can be expressed
as $D_{m\,p}^j(\varphi,\theta,0)=\exp(-i m
\varphi)d_{m\,p}^j(\theta)$, where
\begin{eqnarray}
d_{m\,p}^j(\theta)&=&\left[ (j+p)! (j-p)! (j+m)! (j-m)!
\right]^{1/2}
\times \nonumber \\
&& \sum_s \frac{(-)^s}{(j-m-s)! (j+p-s)! (s+m-p)! s!} \times
\nonumber \\
&& \left(\cos \frac{\theta}{2}\right)^{2j+p-m-2s} \left(-\sin
\frac{\theta}{2}\right)^{m-p+2s}. \label{reducedrot}
\end{eqnarray}
If we put together Eqs. (\ref{rotation}),(\ref{pwaverot}) and
(\ref{reducedrot}), we obtain
\begin{eqnarray}
&& {\mathbf A}= \sum_{j=1}^{\infty}\sum_{m=-j}^j i^j (2j+1)^{1/2}
\int_0^\pi\sin(\theta)d\theta d_{m\,p}^j(\theta) \nonumber \\
&&\int_0^{2\pi}d\varphi \exp(-i m \varphi) g(\theta,\varphi) \left[
{\mathbf A}_{jm}^{(m)} + i {\mathbf A}_{jm}^{(e)}\right].
\end{eqnarray}

Now, I will use the cylindrically symmetric function $g$ from Eq.
(\ref{changevar}) to perform the integrals. In this way, the
$\varphi$ integral is immediate and the following result holds
\begin{eqnarray}
{\mathbf A}&=& 2\pi \sum_{j=\|l+p\|}^{\infty} i^j (2j+1)^{1/2} C_{j
l p} \left[ {\mathbf A}_{j\,(l+p)}^{(m)} + i {\mathbf
A}_{j\,(l+p)}^{(e)}\right]
\nonumber \\
C_{j l p} &=& \int_0^\pi\sin(\theta)d\theta\,
d_{(l+p)\,p}^j(\theta)\, f_l(k sin(\theta)). \label{general}
\end{eqnarray}
This equation is the main result of this letter. Note that Eq.
(\ref{general}) is a valid solution of the Maxwell equations, as
$f_l(k_r)$ is not restricted to paraxial beams, and then can be
applied to a wide number of situations. Also, note that all the
multipolar beams in the superposition share the same value of
$m=l+p$. This has the obvious meaning of summing the OAM and SAM in
the paraxial approximation, but in the more general case, simply
implies that we have been able to form a family of fields with a
well defined $J_z$ value. This solution is different from that
presented in Refs. \cite{allen-book,barnett02,BarnettAllen94} and it
does not have to fulfil the angular momentum equations derived
there.

On the other hand, this result has an interesting side effect. As
multipole fields only depend on the wavelength, we can control the
multipolar content of a field with only paraxial fields. I will step
back a little bit and assume the paraxial approximation again. In
this case, the amplitude of the multipolar field of order
$(j,m=l+p)$ can be approximated by
\begin{eqnarray}
C_{j\; l\, p} = (-)^{l} \left[ (j+p)! (j-p)! (j+l+p)! (j-l-p)!
\right]^{1/2} \nonumber \\
\sum_s \frac{(-)^s}{(j-l-p-s)! (j+p-s)! (s+l)! s!}
\frac{M_{l+2s}\{f_l\}}{2^{l+2s}} \label{amplitude}
\end{eqnarray}
where the number $M_{a}\{f_l\}=\int_0^\infty k_r \, dk_r k_r^a
f_l(k_r)$ is the momentum of order $a$ of the function $f_l$.

Let me now discuss an important class of functions, the
Laguerre-Gaussian modes (LG). The LG modes are defined by two
indices, the OAM index, which I conveniently call $l$ and an index
stating the number of transversal nodes of the function, which I
call $q$. Then, the $f_l$ functions read
$f_{l,q}(k_r)=\sqrt{\frac{w_0^2 q!}{2\pi
(|l|+q)!}}(w_0k_r/\sqrt{2})^{|l|}\, L_q^{|l|}(w_0^2k_r^2/2)
\exp(w_0^2k_r^2/4+i(q-|l|/2)\pi)$, where $w_0$ is the beam width of
the LG mode in real space, and $L_q^l(x)$ are Laguerre polynomials.
When this expression is used, the set of momenta $M_{a}$ can be
calculated analytically and for the special case of LG modes with
$q=0$, we can write the sum in Eq.(\ref{amplitude}) in a strikingly
compact form
\begin{eqnarray}
&&C_{j;l p}=
\sqrt{\frac{(j+p)!(j-p)!(j+l+p)!(j-l-p)!}{(j-\frac{|l|-|l+2p|}{2})!}}
 \\
&&k(-)^{j+\frac{l+|l+2p|}{2}}2^{|l|/2+1}w_0^{-2j-1+|l+2p|}L_{j-\frac{|l|+|l+2p|}{2}}^{|l+2p|}\left((w_0/k)^2\right)
\nonumber \label{CforLGp0}
\end{eqnarray}

\begin{figure}
\centering\includegraphics[width=82mm]{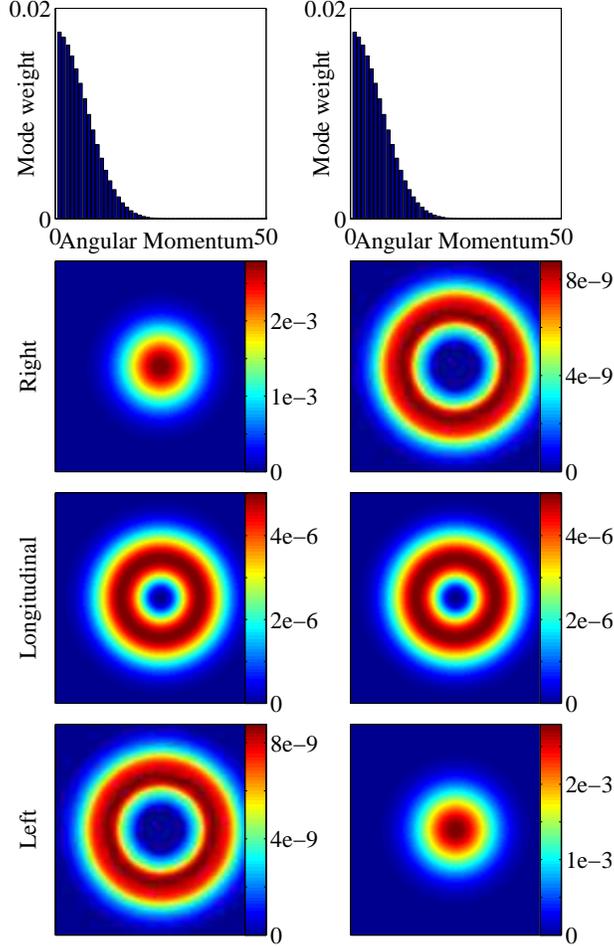} \caption{Multipolar
decomposition of a Gaussian beam. First column: Right circular
polarization ($p=+1$). Second column: Left circular polarization
($p=-1$). First row: weight of the multipolar components
$C_{j\;0\,p}^2$. Second, third and last rows represent the different
components (respectively right circular, longitudinal and left
circular) of the field at $z=0$, after performing the summation of
the multipole components. Note the different scales for the
different components. \label{Gausfield}}
\end{figure}

I will detail now some examples. The simplest case is that of a
paraxial Gaussian beam ($w_0\gg1/k$) with circular polarization. In
this case, formula (\ref{CforLGp0}) gives the following result
\begin{eqnarray}
&&{\mathbf A}= 2\pi \sum_{j=1}^{\infty} i^j (2j+1)^{1/2} C_{j\; 0\, p}
\left[ {\mathbf A}_{j\,p}^{(m)} + i p {\mathbf
A}_{j\,p}^{(e)}\right]
\nonumber \\
&&C_{j\;0 p}=k(-)^{j+1} (j+p)!(j-p)!\sqrt{\frac{1}{(j+\|p\|)!}}
\nonumber \\
&&2w_0^{-2j-1+2\|p\|}L_{j-2\|p\|}^{2\|p\|}\left((w_0/k)^2\right)
\label{CforGAuss}
\end{eqnarray}
Note the following features of this field. First, the field is
composed of multipolar solutions with different total angular
momentum, but with the same projection of the angular momentum in
the direction $z$, which is $+1$ or $-1$ depending on the
polarization of the field. Also, in both cases the amplitudes of the
multipolar components are the same, as $C_{j;0 p}$ does not depend
on the sign of $p$. In Fig. (\ref{Gausfield}) we plot the different
amplitudes of the multipolar components for $w_0=15/k$, and the
field resulting when sum (\ref{CforGAuss}) is numerically performed.

\begin{figure}
\centering\includegraphics[width=80mm]{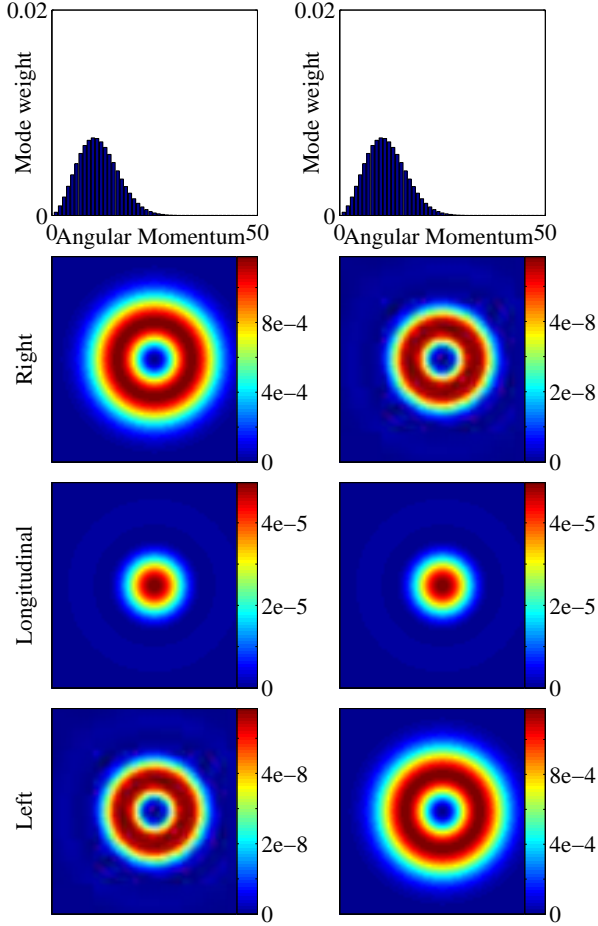} \caption{Multipolar
decomposition of Laguerre Gaussian beams. First column: Right
circular polarization ($p=+1$) and azimuthal index negative
($l=-1$). Second column: Left circular polarization ($p=-1$) and
azimuthal index positive ($l=+1$). First row: weight of the
multipolar components: $C_{j\;\pm1 \, -l}^2$. In this case $C_{j\;+1
\, -1}=-C_{j\;-1 \, +1}$. Second, third and last rows as in Fig.
1.\label{LG1field}}
\end{figure}

Another interesting case occurs  when we use Laguerre-Gaussian beam
with $l=\pm 1$, $q=0$ and circularly polarized $p=-l$. I have shown
that, in the general case, all the multipolar fields have a
component of $J_z$ is equal to $m=l+p$. Then, in this example the
projection of the angular momentum in the $z$ direction of both
considered fields is zero, i.e. $m=l+p=0$. The multipolar
decomposition gives the following result,
\begin{eqnarray}
&&{\mathbf A}= 2\pi \sum_{j=1}^{\infty} i^j (2j+1)^{1/2} C_{j\; \pm
1\, p=-l} \left[ {\mathbf A}_{j\,0}^{(m)} + i p {\mathbf
A}_{j\,0}^{(e)}\right]
\nonumber \\
&&C_{j;l=\pm1 p=-l}=l k(-)^{j} \sqrt{j!(j+p)!(j-p)!} \nonumber \\
&&2^{1/2+1}w_0^{-2j}L_{j-1}^{1}\left((w_0/k)^2\right)
\label{CforLG1}
\end{eqnarray}
 Note that the weights of the multipoles of the two fields have the
same magnitude and opposite signs: $C_{j\; +1\, -1}=-C_{j\; +1\,
-1}$. An example of such fields is found in Fig. (\ref{LG1field}).

This particular example is very interesting as the two considered
fields are decomposed in the same subset of multipolar modes: $
{\mathbf A}_{j\,0}^{(m)}$ and ${\mathbf A}_{j\,0}^{(e)}$. This will
always be the case of fields with opposite polarizations and $\Delta
l=2$. However, in this particular example, both fields share the
same weights in the decomposition. This means that if we produce a
superposition of the kind: ${\mathbf A}=\alpha LG_{1 0} {\mathbf
e_{-1}}+\beta LG_{1 0} {\mathbf e_{-1}}$, we can produce a purely
transversal electric field when $\alpha=-\beta$ or a pure magnetic
one with $\alpha=\beta$. In other cases with opposite polarization
and $\Delta l=2$, one can produce a multipole field with a well
defined parity just in a set of discrete values of $j$, by playing
with the amplitudes of the superpositions and the beam widths of the
two fields.

The examples above, and the formulae (\ref{general}) and
(\ref{CforLGp0}) demonstrate the control of the superpositions of a
set of multipolar fields by using paraxial beams. As considered
above, multipolar fields are the solutions of the electromagnetic
field one should use when treating spherically symmetric problems.
For example, those problems where there is an exchange of angular
momentum  between the electromagnetic field and material particles
are particularly well suited to be treated with multipolar
expansions of the electromagnetic field. One such case are the
electronic transitions in atoms or molecules. It is well known that
multipolar transitions are increasingly more difficult to excite (or
de-excite) for larger $j$'s. This is why almost all the literature
dedicated to the exchange of angular momentum (or more exactly
orbital angular momentum), treats the problem in the dipolar
approximation. Actually an easy calculation shows that the
probability for a multipolar transition to occur scales as
$(a/\lambda)^{2\Delta j}$, with $a$ being the typical radius of the
system involved, and $\Delta j$ the multipolar transition
\cite{Jackson}.

Here we are interested in a different, but related problem. Let's
consider the case where it is needed to control a certain multipolar
transition. This is the case, for example, when our system is in
certain metastable states, as is the case of trapped Ca$^{+}$ ions
in quantum information applications \cite{Reviewjuergen}. In those
cases the rate of transition is low compared to dipole transitions,
simply due to the small size of the ion, compared with the
wavelength of the field. Nevertheless, we can still maximize the
overlap of the laser beam we are using to induce the transition with
the electromagnetic multipole field associated with this transition.
Also, in practical applications we have to deal with paraxial or
near-paraxial beams. In this situations we can use Eq.
(\ref{amplitude}) to engineer the shape and polarization of our
control beam, to maximize the overlap with the desired transition.

Importantly, our method allows to control not only the total angular
momentum exchange involved in a certain transition, but also the
exchange of $J_z$. This can potentially lead to lower losses and
errors in some applications, i.e.  some quantum information
protocols. Our results could be also interesting in the field of
nanophotonics where, in some applications one needs to produce light
which closely resembles the symmetry of a certain system. It is
likely that our method allows to introduce new degrees of freedom in
this kind of applications, considering that our equation
(\ref{general}) is valid beyond the paraxial approximation and as
stated above, in particular can be used with highly focused beams.

In conclusion, our work closes an existing gap between the
descriptions of angular momentum in paraxial and nonparaxial
electromagnetic fields. We have demonstrated that a paraxial beam
with OAM $l$ and polarization $p$ can be decomposed into multipolar
modes with a fixed component of the angular momentum $m=l+p$ and
different components of the total angular momentum $j$. We have
provided analytical rules to calculate the weight of the
superpositions for several interesting cases, thus allowing a
control over the superposition of the multipolar fields. These rules
could be of utility in fields where the interaction of light and
matter has to be controlled with precision.




\end{document}